\documentclass[12pt]{article}
\usepackage{hyperref}
\usepackage{lscape}
\usepackage{amsmath}
\usepackage{amssymb}

\renewcommand{\title}[1]{%
    \bigskip%
    \begin{center}%
    \Large\bf #1%
    \end{center}%
    \vskip .2in}

\renewcommand{\author}[1]{%
    {\begin{center}
    #1
    \end{center}}}
\newcommand{\address}[1]{\vspace{-1.7em}\vspace{0pt}
    {\begin{center}
    \it #1
    \end{center}}}

\begin{document}


\title{Taming galileons in curved spacetime}

\author
{
Rabin Banerjee  $\,^{\rm a,b}$,
Pradip Mukherjee $\,^{\rm c,d}$}
\address{$^{\rm a}$S. N. Bose National Centre 
for Basic Sciences, JD Block, Sector III, Salt Lake City, Kolkata -700 098, India }

\address{$^{\rm c}$Department of Physics, Barasat Government College,\\10, KNC Road, Barasat, Kolkata - 700124.

 }

\address{$^{\rm b}$\tt rabin@bose.res.in}
\address{$^{\rm d}$\tt mukhpradip@gmail.com}

\begin{abstract}
Localising the galileon symmetry along with Poincare symmetry we have found a version of galileon model coupled with curved spacetime which retains the internal galileon symmetry in covariant form. Also, the model has second order equations of motion.
\end{abstract}

\section{Introduction}
The effective field theory approach has provided new insights and results in various contexts. Since these theories are defined by appropriate symmetries, new possibilities are always being suggested and probed. Over the last few years the viable symmetries of systems consisting of one or more scalar fields coupled to gravity have been explored due to their relevance in cosmological model building. This work was triggered by the introduction of a scalar field theory, called Galileon \cite{N}. It was obtained by taking the decoupling limit of the DGP model \cite{D} leading to an effective scalar theory argued to describe the scalar sector of the original model \cite{LP, NR} \footnote{The galileon field is a scalar under Poincare transfprmations. However, unlike the usual scalar fields, it has}.

Galileon theories are interesting in several ways. They are characterised by two properties.
Despite being higher derivative theories, they are free of Ostogradsky-type ghosts since they have no more than second order equations of motion. Also, they have a new nonlinear internal symmetry, called shift symmetry. A direct consequence of this invariance is that the coefficients of the leading galileon interactions are, perturbatively, not renormalised \cite{nima}.

Despite its attractive features, the coupling of galileons to gravity is a tricky and challenging issue. A simple minimal coupling leads to a breakdown of the first property. It is possible to introduce a nonminimal coupling \cite{V} which, however, does not retain the second property. In any case the global shift symmetry transformation , as it stands, is meaningless in curved space. This has resulted in a general scheme for coupling scalar theories to gravity, incorporating only the first property \cite{DDE}. Attempts to covariantize the shift symmetry and define galileons in curved space have always led to restrictions \cite{GHK, G}, so that the findings are not universal. In fact the general consensus is that it is impossible to couple galileons to gravity while retaining shift symmetry or its curved space extension \cite{P}. This has resulted in the introduction of models where galileon shift symmetry is weakly broken \cite{P}.

In this paper we provide a new approach of coupling galileons to gravity where both properties are incorporated. This approach is based on the concept of localising a global symmetry, an example of which is the shift symmetry. It was introduced earlier \cite{U, HB} to localise the Poincare symmetry to provide an alternative formulation of gravity, called the Poincare gauge theory of gravity. Very recently in a set of papers \cite{MMM1, MMM2, MMM3, MM4, MM5}, we have in collaboration with Mitra, developed a nonrelativistic version of it, the so called galilean gauge theory, by localising the galilean symmetry. Apart from yielding results related to nonrelativistic spatial diffeomorphism invariance, it gave a dynamical realisation of Newton-Cartan geometry, with or without torsion, which is the basis of nonrelativistic gravity.  By adopting the same approach, we gauge (i.e, localise) both the Poincare symmetry and the shift symmetry to consistently couple galileon scalars to gravity. Interestingly, the new fields introduced to gauge the galileon symmetry, transform in a similar manner as generators of the galileon algebra \cite{GHJT}. This is the main theme of the paper which constitute the whole of section 3. For convenience of the reader we start with a brief review of Poincare gauge theory in section 2. Finally we conclude in section 4.
\section{Poincare gauge theory -- a review}
  We review the algorithm of constructing Poincare gauge theory (PGT) \cite{ U, Kibble:1961ba,Blagojevic:2002du}. Consider a field theory with action
 \begin{equation}
S = \int d^4x {\cal{{L}}}(\phi,\partial_{k}\phi)\label{gaction}
\end{equation} 
 invariant under 
   the global Poincare transformations in the Minkowski space:
\begin{equation}
x^\mu \rightarrow x^\mu + \xi^\mu\label{GPT}
\end{equation}
where 
$$\xi^\mu = \theta^\mu_{\ \nu}x^\nu + \epsilon^\mu$$
 with both $\theta^{\mu\nu}$ and $\epsilon^\mu$ being infinitesmal constants and $\theta^{\mu\nu}$ antisymmetric. The purpose of PGT is to build an appropriate theory which is symmetric under the local version of (\ref{GPT}). i.e. when the parameters of transformation vary from point to point. Clearly the corresponding transformation is locally Poincare. We construct a local basis ${\bf{e}}_i$  at each spacetime point which are related to the coordinate basis ${\bf{e}}_\mu$ by 
\begin{equation}
{\bf{e}}_i = \delta^\mu_i{\bf{e}}_\mu.\label{basis}
\end{equation}
The fields will be assumed to refer to the local basis. The Latin indices refer to the local Lorentz frame and the Greek indices refer to the coordinate frame. The differentiation between the local and the coordinate basis appears to be a formal one here but it will be a necessity in curved space -- time.

   Now
under (\ref{GPT}) the action (\ref{gaction}) transforms as $$S \to S + \Delta S$$ where
\begin{equation}
\Delta S = \int d^4x \Delta {\cal{{L}}}(\phi,\partial_{k}\phi) \label{gactionvar}
\end{equation}
The variation
 $\Delta{\cal{L}}$ is given by
\begin{equation}
\Delta{\cal{L}} = \delta{\cal{L}} + \xi^\mu\partial_\mu{\cal{L}}+ \partial_\mu\xi^\mu{\cal{L}}
\label{oinvariance}
\end{equation}
where $\delta{\cal{L}}$
 is the form variation of the lagrangean
\begin{equation}
\delta{\cal{L}} = {\cal{L^\prime}}(x) - {\cal{L}}(x)\label{fv}
\end{equation}
${\cal{L^\prime}}$ being the transformed lagrangean {\footnote{The form variation of a quantity will always be denoted by the precedent $\delta$.}}. The condition for invariance of the theory is,
\begin{equation}
\Delta{\cal{L}} = 0
\label{invariance}
\end{equation}

  It is useful to scrutinise the above invariance condition with care. 
For global Poincare transformations 
\begin{equation}
\partial_\mu\xi^\mu=0\label{condition1}
\end{equation}
Also 
the field and its derivatives transform as
\begin{eqnarray}
\delta\phi &=& \left(\frac{1}{2}\theta^{ij}\Sigma_{ij} - \xi^\mu\partial_\mu\right) \phi={\cal{P}}\phi\label{condition2a}\\
\delta\partial_k\phi &=& \left(\frac{1}{2}\theta^{ij}\Sigma_{ij} - \xi^\mu\partial_\mu\right)\partial_k\phi + \theta_k^{\ j}\partial_j\phi \nonumber\\
&=& {\cal{P}}\partial_k\phi + \theta_k^{\ j}\partial_j\phi\label{condition2b}
\end{eqnarray}
$\Sigma_{ij}$ are the Lorentz spin matrices, the form of which depend on the particular representation to which $\phi$ belong. These are matrices with constant elements that satisfy the Lorentz algebra
\begin{equation}
\left[\Sigma_{ij},\Sigma_{kl}\right] = \eta_{il}\Sigma_{jk} -
      \eta_{ik}\Sigma_{jl} + \eta_{jk} \Sigma_{il} - \eta_{jl}\Sigma_{ik}\label{algebra}  
\end{equation}
Equations (\ref{condition1}), (\ref{condition2a}) and (\ref{condition2b}) are instrumental for the invariance condition (\ref{invariance}) to be satisfied.  The form variation $\delta{\cal{L}}$ is explicitly given by
\begin{equation}
\label{lvariation}
\delta{\cal{L}}= \frac{\delta {\cal{L}}}{\delta\phi}\delta\phi + \frac{\delta {\cal{L}}}{\delta\partial_k\phi}\delta\partial_k\phi
\end{equation}
and it is the form of variations given by (\ref{condition2a}), (\ref{condition2b})and the condition (\ref{condition1}) which lead to the invariance of the theory. 

  Advantage of the local basis is that in it the  fields transform formally in the same way as in global transformation
 \begin{eqnarray}
\delta\phi &=& \left(\frac{1}{2}\theta^{ij}\Sigma_{ij} - \xi^\mu\partial_\mu\right) \phi\nonumber
           ={\cal{P}}\phi
\end{eqnarray}
But 
  their derivatives $\partial_k\phi$ transform as
\begin{eqnarray}
\delta\partial_k\phi &=& \left(\frac{1}{2}\theta^{ij}\Sigma_{ij} - \xi^\mu\partial_\mu\right)\partial_k\phi - \partial_k\xi^\nu\partial_\nu\phi + \frac{1}{2}\partial_k\theta^{ij}\Sigma_{ij}\phi\nonumber\\
           &=& {\cal{P}}\partial_k\phi - \partial_k\xi^\nu\partial_\nu\phi + \frac{1}{2}\partial_k\theta^{ij}\Sigma_{ij}\phi\label{condnew} 
\end{eqnarray}
which is different from their counterpart in equation (\ref{condition2b}).
  When the Poincare symmetry (\ref{GPT}) is assumed to be a local symmetry the parameters $\theta$ and $\epsilon$ are no longer constants but becomes function of space -- time coordinates.  
 Here, it is advantageous to take the  parameters  $\xi^\mu = \theta^\mu_{\ \,\nu} x^\nu +  \epsilon^\mu$ and $\theta^{ij}$ as the independent parameters. It is natural that the action (\ref{gaction}) which was invariant under (\ref{GPT}) with constant parameters ( global Poincare transformations ) will cease to remain invariant when the parameters become function of spacetime ( local Poincare transformations ). Also, equation (\ref{condition1}) is no longer true.

 To achieve invariance of the matter action under the {\it{local}} Poincare transformations, the above departures must be corrected. The first thing is to replace the ordinary derivative $\partial_k\phi$ by some covariant derivative $\nabla_k\phi$ which will transform  as in (\ref{condition2b}).
 This is done in two steps:

(i) In the first step the $\theta$ - covariant derivative $\nabla_\mu$ is introduced which eliminates the $\partial_\mu\theta^{ij}$ term from (\ref{condnew}). We define  $\nabla_\mu$ as
\begin{equation}
\nabla_\mu = \partial_{\mu} + \frac{1}{2}\omega^{ij}_{\ \ \mu}\Sigma_{ij}\label{nablamu}
\end{equation}
where $\omega^{ij}_{\ \ \mu}$ are the `gauge potentials'.The required transformation of $\nabla_\mu\phi$ is read off from (\ref{condnew}) as
\begin{equation}
\delta\nabla_\mu\phi = {\cal{P}}\nabla_{\mu}\phi -\partial_\mu \xi^{\nu}\nabla_\nu\phi
\end{equation}
The transformation of the 'gauge field' 
$\omega^{ij}_{\ \ \mu}$ is determined from this requirement.

(ii) In the second step the covariant derivative in the local frame is constructed as 
\begin{equation}
\nabla_k = b_k^{\ \mu}\nabla_\mu\label{covder}
\end{equation}
where $b_k^{\ \mu}$ is another compensating field. For later convenience we define
$b^k_{\ \mu}$ as the inverse to $b_k^{\ \mu}$. 
If the fields $b^i_{\ \mu}$ and $\omega^{ij}_{\ \ \mu}$ transform as 
\begin{eqnarray}
\delta b^i_{\ \mu} &=& \theta^i_{\ k} b^k_{\ \mu} - \partial_\mu\xi^\rho b^i_{\ \rho} - \xi^{\rho}\partial_{\rho} b^i_{\ \mu}\nonumber\\
\delta \omega^{ij}_{\ \ \mu} &=& \theta^i_{\ k} \omega^{kj}_{\ \ \mu} + \theta^j_{\ k} \omega^{ik}_{\ \ \mu} - \partial_\mu\theta^{ij} - \partial_\mu\xi^\rho \omega^{ij}_{\ \ \rho}\nonumber
\\
 &-& \xi^{\rho}\partial_{\rho}\omega^{ij}_{\ \ \mu}\label{fieldtrans}
\end{eqnarray} 
then  $\nabla_k\phi$
 transforms as 
\begin{equation}
\delta\nabla_k\phi = {\cal{P}}\nabla_k\phi + \theta_k^{\ j}\nabla_j\phi\label{C}
\end{equation}
This transformation rule is formally identical with
 (\ref{condition2b}).
  The matter lagrangian density ${\cal{L}} = {\cal{L}}(\phi,\partial_{k}\phi)$ which was invariant under global Poincare transformations is converted to an invariant density ${\cal{\tilde{L}}}$ under local Poincare transformations by replacing the ordinary derivative $\partial_{k}$ by the covariant derivative $\nabla_k $ i.e. 
$${\cal{\tilde{L}}} = {\cal{\tilde{L}}}(\phi,\nabla_{k}\phi).$$ 
The departure from equation (\ref{condition1}) can be accounted for by altering the measure of spacetime integration suitably. An invariant action is now constructed as 
\begin{equation} S = \int d^4x b{\cal{\tilde{L}}}(\phi,\nabla_{k}\phi)\label{m}
\end{equation}
 where $b = {\rm det}\,b^i_{\ \mu}.$ The invariance of this action is ensured by the transformations (\ref{fieldtrans}) of the 'potentials'.

      At this point a very interesting property of the above construction should be noted. The transformations (\ref{fieldtrans}) comprise the Poincare gauge transformations. Their structure suggests a geometric interpretation. The basic fields $b^i_{\ \mu}$ and $\omega^{ij}_{\ \ \mu}$ mimic the tetrad and the spin connection in curved spacetime. The most general invariance group in curved spacetime consists of the LLT plus diff. Observe in (\ref{fieldtrans}), that the Latin indices transform as under LLT with parameters $\theta^{ij}$, and the Greek indices transform as under diff with parameters $\xi^\mu$. This suggests a correspondence between the Poincare gauge transformations with the geometric transformations of the curved spacetime. 
       The geometric interpretation of PGT is a crucial step. It emerges from the application of the gauge principle. Here it is useful to mention that passage to flat limit of the theory (\ref{m}) is smooth. The original theory (1) is recovered simply by putting all new fields zero and making the transformation parameters space time independent.

\section{Construction of the covariant galileon model}
In four dimensions, there are five different parts of the galileon model ${\cal{L}}_i, i=1,2..,5$, which are complete in themselves. ${\cal{L}}_1=\pi$ is trivial and ${\cal{L}}_2 = \partial_\mu\pi\partial^\mu\pi$ is the usual quintessence. The first nontrivial Gallileon term is
\begin{eqnarray}
{\cal{L}}_3 = \Box{\pi}
\left(\partial\pi\right)^2\label{L3}
\end{eqnarray} 
As has been mentioned in the introduction, all the nontrivial galileon parts satisfy the following properties: 
\begin{enumerate}
\item Despite being higher derivative theories, they are free of Ostogradsky-type ghosts since they have no more than second order equations of motion.
\item They are invariant under a nonlinear symmetry transformation of the scalar field $\pi$,
\begin{eqnarray}
 \pi \to \pi + c + b_\mu x^\mu \label{shift}
\end{eqnarray}
\end{enumerate} 
where $c$ and $b_\mu$ are constants.

We want to couple (\ref{L3}) with gravity. A la PGT, a review of which os provided in section 2,       
the problem may be posed in the following way. We have a theory in the flat space with the generic action, 
\begin{equation}
\int d^4x{\cal{L}}\left(\pi,\partial_\mu \pi, \partial_\mu\partial_{\nu}\pi
\right)\label{action}
\end{equation}
 which is invariant under the combined Poincare and Galileon transformations,
\begin{eqnarray}
\delta x_{\mu}&=& \xi_{\mu}\nonumber\\
\delta\pi &=& -\xi^\lambda\partial_\lambda \pi+ c + b_\mu x^\mu \nonumber\\
\delta\partial_\mu\pi &=& -\xi^\lambda\partial_\lambda
\partial_\mu \pi + \theta_\mu{}^
\lambda\partial_\lambda\pi +b_\mu\nonumber\\
\delta\partial_\mu\partial_\nu\pi &=& -\xi^\lambda\partial_\lambda
\partial_\mu\partial_\nu \pi + \theta_\mu{}^
\lambda\partial_\lambda
\partial_\nu\pi+ \theta_\nu{}^
\lambda\partial_\mu
\partial_\lambda\pi \label{flat}
\end{eqnarray}
Here, $\xi^\lambda = \epsilon^\lambda + \theta^\lambda{}_\rho x^\rho$ are the infinitesimal Poincare transformation parameters. What are the modifications to be done in (\ref{action}) and (\ref{flat}) when gravity is included? To find the answer  is the principal aim of the paper.

The algorithm already stated requires a gauging of the original (global) symmetry. We wish to construct an action, starting from (\ref{action}), that is invariant under the local version of (\ref{flat}). By local version of the transformations, we mean that the transformation parameters $\epsilon^\mu,\theta^\mu{}_\nu, c, b_\mu$ varies from point to point. The transformation 
\begin{equation}
\delta \pi=-\xi^{\lambda}(x)\partial_{\lambda}\pi(x)+c(x)+b_{\mu}(x)x^{\mu}\label{A1}
\end{equation}
is not galileon (plus
Poincare) transformations in the global sense. But in terms of the local Lorentz frame, it plays the role of galileon transformation in the neighbourhood of the origin. However,
there is an obvious nontrivial change in the transformation of the derivatives, e.g.,
\begin{equation}
\delta\partial_\mu\pi = -\xi^\lambda\partial_\lambda
\partial_\mu \pi-(\partial_{\mu}\epsilon^{\lambda}+(\partial_{\mu}{\theta^{\lambda}}_{\rho})x^{\rho})\partial_{\lambda}\pi +\partial_{\mu}c+ \theta_\mu{}^
\lambda(x)\partial_\lambda\pi +b_\mu(x)+(\partial_{\mu}b_{\sigma})x^{\sigma}\label{A2}
\end{equation} is the form variation of $\partial_\mu\pi$ defined as $\cal{L}$ in (\ref{fv}), which disturbs the symmetry of (\ref{action}) under
(\ref{A1}). Our method is to devise appropriate changes to 
(\ref{action} so that the symmetry is preserved. Correspondence to the global symmetry can be
easily established --the prescription is to remove the spacetime dependence of the parameters exactly following PGT.

Coming back to the original flat space models, the Galileon transformations have already been defined in (\ref{shift}). 
 The symmetry under (\ref{flat})is due to the structure of the transformations of $\pi$,  $\partial_\mu\pi$
 and $\partial_\mu\partial_\nu\pi$. Together they ensure that the total change of ${\cal{L}}$ is given by \footnote{For the sector involving only the Poincare symmetry, $\Delta {\cal{L}} =0$.},
\begin{eqnarray}
\Delta {\cal{L}} = \delta {\cal{L}} + \xi^\lambda\partial_\lambda{\cal{L}} +\partial_ \lambda \xi^\lambda
{\cal{L}} = 2\partial^\lambda\left[	\left(\partial_\lambda\pi\partial_\mu\pi - \frac{1}{2}\eta_{\mu\lambda}\left(\partial_\alpha\pi\partial^\alpha\pi\right)\right)\right]b^\mu\label{cond}
\end{eqnarray} 
where, $\delta {\cal{L}}$ is the form variation of ${\cal{L}}$.
 
We observe that the invariance is at the quasi level because $\Delta {\cal{L}}$ does not vanish but changes by a partial derivative. This quasi invariance is due to the galileon invariance and rests on 
\begin{eqnarray}
\partial_\mu b_\nu = 0 \label{cond1}
\end{eqnarray}
In order to construct an action that is invariant under the local symmetry (\ref{A1}, \ref{A2}) we have to introduce a local coordinate system (as mentioned above) as in (\ref{basis}).
   
   Our task is then clear. We have to replace $\partial_\mu\pi$ and $\partial_\mu\partial_\nu\pi$
   by their local counterparts 
    such that they transform form
   invariantly as in (\ref{flat}).
  In the more familiar context of Poincare invariance only, the covariant derivatives $\nabla_a\pi$ and $\nabla_a\nabla_b\pi$  are defined in the following form \cite{HB, M1}
  \begin{eqnarray}
 \nabla_a\pi = \Sigma_a{}^\mu D_\mu \pi\hskip.1cm ;\hskip.1cm D_\mu \pi = \partial_\mu \pi + \frac{1}{2}B_\mu{}^{ab}\sigma_{ab}\pi
 \label{cov1}\\
 \nabla_a \nabla_b \pi = \Sigma_a{}^\mu D_\mu \left(\Sigma_a{}^\mu D_\mu \pi\right)\label{cov2})
\end{eqnarray} 
where $\sigma_{ab}$ is the Lorentz spin matrix whose form is dictated by the spin of the field $\pi$.  Here $\Sigma_a{}^\mu$ and $B_\mu{}^{ab}$ are new gauge fields corresponding to translation and lorentz transformation (see section 2)\footnote{The gauge fields corresponding to the Lorentz rotation do not appear because $\pi$ is a Lorentz scalar. }. The galileon field has an additional shift symmetry. To compensate this, we require to introduce further gauge fields $A_{\mu}$ and $D$. The new covariant derivatives are now defined as,
   \begin{eqnarray}
   {\bar{\nabla}}_a\pi = \Sigma_a{}^\mu{\bar{ D}}_\mu \pi;\hskip .4cm {\bar{ D}}_\mu \pi = \left(D_\mu\pi + F_\mu; F_\mu = A_\mu 
   + x_\mu D\right)\label{10}
   \end{eqnarray}
  The transformations of the new fields are obtained by demanding that the covariant derivatives (\ref{10}) transform as the ordinary one in (\ref{flat}),
 \begin{equation}
 \delta({\bar{\nabla}}_a\pi)=-\xi^c \partial_c({\bar{\nabla}}_a
 \pi)+{\theta_a}^b{\bar{\nabla}}_b\pi
 +b_a
 \end{equation}
 This yields,
 \begin{eqnarray}
 \delta\Sigma_a{}^\mu &=& -\xi^\lambda\partial_\lambda
 \Sigma_a{}^\mu
+\partial_\lambda\xi^\mu
\Sigma_a{}^\lambda + \theta_a
{}^b\Sigma_b{}^\mu\nonumber\\
\delta B_\mu{}^{ab} &=& -\xi^\lambda\partial_\lambda
B_\mu{}^{ab}-\partial_\mu
\theta^{ab}
-\partial_\mu\xi^\lambda B_\mu{}^{ab} + \theta^a{}_c B_\mu{}^{cb} +\theta^b{}_c B_\mu{}^{ac} \nonumber\\
\delta A_\mu &=& -\xi^\lambda\partial_\lambda
A_\mu
-\partial_\mu\xi^\lambda A_\lambda -\partial_\mu c\nonumber\\
\delta \left(
x_\mu D\right) &=& -\xi^\lambda\partial_
\lambda\left(
x_\mu D\right) -\partial_\mu
\xi^\lambda \left(
x_\lambda D\right) -x^\nu\partial_\mu b_\nu \label{trans}
\end{eqnarray}
We observe that $A_{\mu}$ and $x_{\mu}D$ transform as four vectors under local Poincare transformations. Galileon transformations, on the other hand, act like independent translations. The Galileon symmetries manifested through these relations are compatible with analogous results in \cite{GHJT} where the algebra of Galileon generators has been defined.
Also note that 
\begin{eqnarray}
\delta{\bar{D}}_{\mu} \pi & = &-\xi^{\lambda}\partial_{\lambda}({\bar{D}}_{\mu} \pi)-\partial_{\mu}\xi^{\lambda}{\bar{D}}_{\lambda} \pi+b_{\mu}\nonumber\\
\delta F_\mu & = & -\xi^\lambda\partial_\lambda
F_\mu
-\partial_\mu\xi^\lambda F_\lambda -\partial_\mu c - x^\nu\partial_\mu b_\nu \label{trans1}
\end{eqnarray}
Thus both $F_\mu$ and ${\bar{D}}_\mu\pi$ transform as four vectors under local Poincare transformations.

The galileon model (\ref{L3}) also contains second derivative of $\pi$. The transformations of the second derivatives, however, do not have any galileon contribution (see the last eq. of (\ref{flat})). Consequently,
the second derivative $\partial_\mu\partial_\nu\pi$
should be replaced by 
$\nabla_a{\bar{\nabla}}_b\pi$ \footnote{A general discussion of localising the Poincare sector of a higher derivative theory is given in \cite{M1}.}. Explicit calculation shows that under local Poincare plus galileon transformations $\nabla_a{\bar{\nabla}}_b
\pi$ transform in the same way as $\partial_\mu\partial_\nu\pi$ i.e.
\begin{equation}
\delta(\nabla_a {\bar{\nabla}}_b\pi)=-\xi^d \partial_d(\nabla_a {\bar{\nabla}}_b\pi)+{\theta_a}^d (\nabla_d {\bar{\nabla}}_b\pi)+{\theta_b}^d (\nabla_a {\bar{\nabla}}_d\pi)
\end{equation}
provided,
\begin{equation}
\nabla_a b_c = 0\label{cond2}
\end{equation}
This is a natural generalization of the condition (\ref{cond1}) in flat space. 

An important consequence of (\ref{cond2}), to be exploited later, will now be discussed. Defining $F_a$ from $F_{\mu}$ as,
\begin{equation}
F_a={\Sigma_a}^{\mu}F_{\mu}
\end{equation}
and,
\begin{equation}
\nabla_a F_b={\Sigma_a}^{\mu}[\partial_{\mu}({\Sigma_b}^{\nu}F_{\nu})+B_{\mu}^{bc}{\Sigma_c}^{\nu}F_{\nu}]
\end{equation} 
we find,
\begin{equation}
\nabla_a F_b-\nabla_b F_a={\Sigma_a}^{\mu}{\Sigma_b}^{\nu}(\partial_{\mu}F_{\nu}-\partial_{\nu}F_{\mu})\label{f}
\end{equation}
for zero torsion, 
\begin{eqnarray}
\nabla_a{\Sigma_c}^\mu - \nabla_c{\Sigma_a}^\mu = 0\label{torsion}
\end{eqnarray}
The variation under galileon transformations now yields,
\begin{align}
\delta(\nabla_a F_b-\nabla_b F_a)&={\Sigma_a}^{\mu}{\Sigma_b}^{\nu}(\partial_{\mu}b_{\nu}-\partial_{\nu}b_{\mu})
\notag\\&=(\nabla_a b_b-\nabla_b b_a)=0
\end{align}
on account of (\ref{cond2}). This implies that the choice, 
\begin{equation}
(\nabla_a F_b-\nabla_b F_a)=0 \label{x}
\end{equation}
may be consistently implemented. 

When the transformations are localised the theory invariant under local transformations is obtained by replacing the first order and second order derivatives by the corresponding covariant derivatives. Coming back to the galileon action, we write its localised form as
\begin{equation}
S =\int d^4x \frac{1}{\Sigma}\bar{{\cal{L}}_3}=\int d^4x \frac{1}{\Sigma}\left[ (\nabla_a{\bar{\nabla}}^a\pi)(\left({\bar{\nabla}}_b\pi) ({\bar{\nabla}}^b\pi\right)\right]
\label{la}
\end{equation}
where $\Sigma = \det{\Sigma_a}^\mu $ and the factor $\Sigma^{-1}$ corrects the change due to the fact that, contrary to the global case, $\partial_\mu \xi^\mu \ne 0$ (see (\ref{cond})). 

Now it is known that if an action is invariant under the usual (global) transformations, then under local transformations the invariance is recovered by suitably replacing the ordinary derivatives by the covariant derivatives. The same is however not true if the action is only quasi invariant, as happens here. Nevertheless, we proceed with the same algorithm and finally make an explicit check of the invariance.

The theory (\ref{la}) is a theory in the flat space but the transformations of the fields  
suggest a lucrative geometric interpretation. 
The action (\ref{la}) may be viewed as the first order (vielbein) formulation of the galileon model in curved space time. In this interpretation ${\Sigma_a}^\mu$ is the tetrad and $B_\mu{}^{ab}$ are the spin connections \cite{HB}. The situation here is in complete parallel with PGT (see section2). The galileon symmetry being an internal symmetry does not affect spacetime Poincare symmetry. 

 To identify (\ref{la}) as the required galileon model in curved spacetime we have to show that the properties 1 and 2 hold. It is easy to show that the higher (third) order derivatives cancel in the corresponding Euler Lagrange equations. Hence (\ref{la}) is ghost free.
To prove that the shift symmetry holds, we first calculate the variation of $\bar{\cal{L}}_3$ from (\ref{la}), 
\begin{equation}
\delta{\bar{{\cal{L}}_3}} =
2\nabla_a\left[{\bar{\nabla}
}^a\pi{\bar{\nabla}
}_c\pi b^c - \frac{1}{2}\delta_c^a\left({\bar
{\nabla}
}\pi\right)^2b^c\right]-2({\bar{\nabla}
}^a\pi)b^c(\nabla_a F_c- \nabla_c F_a)
\label{ss}
\end{equation}
Using the condition (\ref{x}), the quasi invariance of the galileon action (\ref{la}) is proved.

A very interesting aspect of our result (\ref{la}) will be appropriate to mention before conclusion. From (\ref{f}) and (\ref{x}) we can write
$F_\mu$ as a pure gauge,
 \begin{equation}
F_\mu = \partial_\mu\phi 
\end{equation}
for some scalar field $\phi$. Using this one obtains
\begin{equation}
{\bar{D}_\mu}\pi=
\partial_
\mu \pi + F_\mu =
\partial_\mu\left(\pi + \phi\right)= \partial_\mu \pi^\prime
\end{equation}
with $\pi^\prime$ is a new field. The curved space galileon model (\ref{la}) can entirely be written in terms of $\pi^\prime$. The parallel with Stuckelberg construction in the context of Proca model is evident. However, in this case, the 
Stuckelberg field is an outcome of our analysis.
\section{Conclusion}
To conclude, our goal has been achieved. By following a systematic algorithm of localising symmetries \cite{U,Kibble:1961ba,HB, Blagojevic:2002du, MMM1, MMM2, MMM3, MM4, MM5} we have gauged the original (global) Poincare and galileon symmetries of the model (\ref{L3}). This approach was pioneered by \cite{U, Kibble:1961ba} in constructing thr Poincare gauge theory (PGT), a brief review of which has been provided. A notable characteristic of PGT is that , it allows a geometric interpretation so that the 
localised theory in flat space can be considered as the original theory coupled with gravity. This gave us an opportunity to construct curved galileon theory which is an open problem,

 In the present paper we have localised the Poincare and Galileon symmetrits of the usual galileon model (\ref{L3}) simultaneously. This led to the construction of a new model (\ref{la}). Remarkably,(\ref{la}) does not lead to any higher order equations of motion and is also (quasi) invariant under (\ref{A1}), which is interpreted as the appropriate covariant version of the galileon symmetry (\ref{flat}). Following the results of PGT this is interpreted as the galileon model (\ref{L3}) coupled with gravity.  The passage to the flat limit is smooth and is simply obtained by setting the new fields $({\Sigma_a}^{\mu}, B_{\mu}{}^{ab}, A_{\mu}$ and $D)$ to zero. Although explicit results were given for ${\cal{L}}_3$ (\ref{L3}), the method is generic and easily applicable to other galileon invariant models. 

\end{document}